# Building Random, Fair, and Verifiable Games on Blockchain. Raffle smart contract designs on Sui Network

Eason Chen
Carnegie Mellon University

Justa Liang
Bucket Protocol

Ray Huang
Bucket Protocol

Pierce Hung
Bucket Protocol

Damien Chen
Bucket Protocol

Ashley Hsu
Bucket Protocol

Konstantinos Chalkias
Mysten Labs

Stefanos Pleros
Mysten Labs

## ABSTRACT

Randomness plays a pivotal role in modern online gaming. However, there have been controversies that the claimed winning probabilities in gaming do not align with reality, leading gaming companies to face legal challenges and even financial setbacks. Fortunately, blockchain-based games offer a solution to the transparency and fairness issue regarding randomness. Furthermore, emerging blockchain technology like Sui Network enhances the efficiency of smart contracts by eliminating traditional web3 barriers, such as inefficiencies and expensive transaction fees. This unlocks the potential for extensive decentralized gaming applications.

This paper aims to provide insights into designing a fair, verifiable, and efficient smart contract game on blockchain by the example of building raffles on the Sui Network. We explore efficient methods for implementing randomness on smart contracts, including DRAND committee-based decentralized random beacons and single private-key-based verifiable random functions (VRF). Subsequently, progress from basic to comprehensive smart contract design. By taking advantage of smart-contract native cryptographic primitives, we addressed limitations in developing blockchain games in general, such as data input and storage space constraints.

We propose corresponding solutions, encompassing the utilization of Object Tables, Delegate Object Creation, and Zero-Knowledge Proofs (ZKP) to optimize storage and input efficiency. After testing our designs, we found that the transaction fees for DRAND beacons and private-key-based house-owned VRFs are similar. Moreover, Object Tables incur higher overall transaction fees, while the ZKP setup fee is very cheap but becomes very expensive during the verification process. Moreover, we identified suitable designs for different application scenarios by comparing the pros and cons of different smart contract implementations. Our findings provide valuable guidance for future researchers and developers in building random, fair, and verifiable games with smart contracts, with a focus on gas optimization and security among the others.

## 1 INTRODUCTION

Randomness is a crucial element within games [18]. These chance-driven rewards introduce thrill and anticipation, enticing players to engage and potentially win valuable prizes, attracting many players willing to spend money on it. In the game industry, the player count is projected to reach 3.38 billion in 2023, generating a revenue of $187.7 billion with a promising upward trajectory [17]. Games with randomized raffles and gacha mechanics contribute significantly to the game industry revenue stream because they encourage players to spend money until they get the desired prizes.

Nevertheless, there were doubts raised regarding the fairness of randomness among games. For example, in 2020, the game "Lineage M." advertised a 5% probability of obtaining a rare virtual item through a raffle. After spending $129,000 in the raffle, Taiwanese streamer Ding suspected the true probability was only around 2.3%. Subsequently, he filed a lawsuit against the operating company of "Lineage M." As a result of the lawsuit, the company had to pay Ding approximately $243,000 in compensation and also faced a government-imposed fine of approximately $60,000 [14]. Consequently, this prominent gaming company, which held Taiwan's second-largest gaming market value, suffered significant damage to its reputation and stock price with a two-day consecutive limit down. Six months after the lawsuit, the company's stock price had plummeted by 30%, reducing its market value by $1.4 million. Unfortunately, despite such painful lessons from the past, most games continue to conceal the true probabilities of their raffle.

Fortunately, players and game developers have started recognizing the importance of transparent randomness. One of the most promising methods to achieve such transparency is by leveraging web3 smart contracts in game development. These contracts ensure that winning probabilities are publicly visible, immune to tampering, and easily verifiable. Additionally, blockchains also guarantee that players have full ownership and transaction rights over in-game assets. Due to these reasons, games built on blockchain have experienced substantial growth in recent years. As of 2023, the product category on web3 with the most active users is no longer DeFi but Games [12].

Among these web3 technologies, the newly introduced Sui Network shows significant promise due to its low latency and transaction fees, cryptographic primitives availability and high scalability, making it a favored choice among many game developers. However, it is crucial to highlight that integrating randomness and games into the blockchain like Sui involves several critical considerations. Neglecting the above can result in cost and programmability efficiencies and potential system crashes.

At present, despite the emergence of gambling-centric smart contract initiatives, there remains a noticeable void in scholarly and technical literature that comprehensively addresses the nuances of instituting randomness-driven games on modern high-caliber blockchains. In light of this, our manuscript seeks to serve as a beacon for forthcoming blockchain game creators, using the development of a random, equitable, and auditable raffle game on Sui as a case study. Our primary research inquiry is as follows:

- How can one construct a fair, verifiable, and cost-efficient raffle game employing smart contracts within the Sui Network?

## 1.1 Paper organization

In this paper, we will first review the background of web3 technologies and explain why the Sui is more suitable for game development than traditional blockchains. Next, we will describe several methods for implementing blockchain randomness, discussing advantages and limitations. Then, by progressing from simple to comprehensive designs, we elucidate the process of implementing a random raffle with fair, verifiable, and efficient smart contract designs on the Sui Network. We will discuss the advantages and limitations of each design and explore strategies to address these limitations. Finally, we will summarize the content above and provide an overall discussion and comparison of different designs.

## 2 BACKGROUND

### 2.1 Blockchain and Smart Contract

Blockchain is a distributed ledger technology (DLT) that securely records transactions across a network, ensuring transparency, immutability, and data trust [25]. Smart contracts are code-based logic running on the blockchain, capable of executing specific actions when predefined conditions are met, thereby automating a multitude of processes. These contracts function within blockchain networks, and presently, numerous decentralized applications (DApps) have been developed across various blockchains utilizing smart contracts.

Every call to a blockchain function, such as writing data or executing smart contracts, requires costs known as transaction fees [6]. This practice is indispensable due to the decentralized nature of the blockchain, where computations and data storage occur across multiple nodes. Additionally, to ensure the consistency, transactions must be executed sequentially and packaged into blocks, which can limit the computational capacity of the platform. Transaction fees serve as a means to prevent resource misuse and maintain cost equilibrium.

However, high transaction fees represent a significant obstacle to the widespread adoption of blockchain applications, as users may be disincentivized from using DApps when transaction fees surpass their potential earnings. Consequently, DApps about earning, such as Finance, Non-Fungible Tokens (NFT), and Gaming Finance, currently dominate the landscape, whereas those emphasizing playability attract fewer users [12].

Thankfully, several modern groundbreaking DLT designs, such as the Sui Network, have emerged to reduce the computational and storage burdens on the network while enhancing computational efficiency, thus alleviating transaction fee issues.

### 2.2 Sui Network

Mainnet launched in May 2023. Sui is a decentralized, permissionless smart contract platform prioritizes low-latency asset management [23]. Originating from Meta's Diem (formally known as Libra), Sui utilizes the Move programming language [1] to define and manage assets owned by addresses, with custom rules for asset creation, transfer, and mutation. Unlike Ethereum's account-based design, in which assets exist as numerical variables within the address or smart contracts, Sui's asset management is object-based, similar to the Unspent Transaction Output (UTXO) structure found in Bitcoin [16], allowing objects of digital assets to be fragmented, combined, and transferred to different addresses.

Moreover, what sets Sui apart from traditional blockchain networks is its utilization of a Directed Acyclic Graph (DAG) model to record transactions. As shown in Figure 1, each transaction block on Sui includes several transactions with inputs of different objects on Sui Network. Then, these transactions will mutate or create new objects. Using DAG and object-based design, Sui can enable transactions with unrelated objects to be executed without a specific sequence, maximizing Sui's computational efficiency and scalability. As a result, despite Sui experiencing a rapid increase in transaction volume from thousands to tens of millions in a short period, the transaction fees have remained nearly consistent [7].

In addition, Sui has made optimizations in data storage. Sui allows the deletion of data on the network to free up space and obtain a Storage Rebate Fee [23]. For example, in Figure 1, the first transaction requires writing 100 Kilobytes (KB) of participants into an array in the *RaffleObject*. It incurs a computation fee of 0.001 Sui and a storage cost of 0.779 Sui. However, when a subsequent transaction clears the array of participants and thus frees 100 KB of data at that *RaffleObject*, the transaction sender will receive a storage rebate of 0.77 Sui. This ultimately reduces the overall longterm cost of storing a substantial amount of data on Sui, leaving only the computational and necessary storage fees, as extra storage costs can be rebated by clearing data storage after operations are completed.

Notably, even though on-network data is cleared, verification and reproduction remain possible since inputs are stored within the Transaction Block and retained in a database on the Sui Nodes without consuming network hot-memory resources.

## 3 RANDOMNESS PRACTICE IN BLOCKCHAIN

In general, the implementation of verifiable random functions (VRF) on blockchain can be broken down into the following two steps:

(1) **Initiate**: Define validation criteria and operational logic through a smart contract for an unknown and unpredictable random number.
(2) **Verify and Execute**: Once the random value is known, validate it against the specified criteria. If it meets the conditions, execute subsequent logic using this random number.

The following sub-sections explain various practical methods of implementing randomness.

### 3.1 Block-Hash Randomness

In the early days of the Proof-of-Work (PoW) blockchain, the block hash was one method for getting random values. At PoW, miners competed in computational works to be the first to create new valid blocks with hash matching certain criteria [16]. Since the block hash is randomly generated, it can achieve randomness by specifying a hash of a future block in the smart contract logic as the seed of randomness [2].

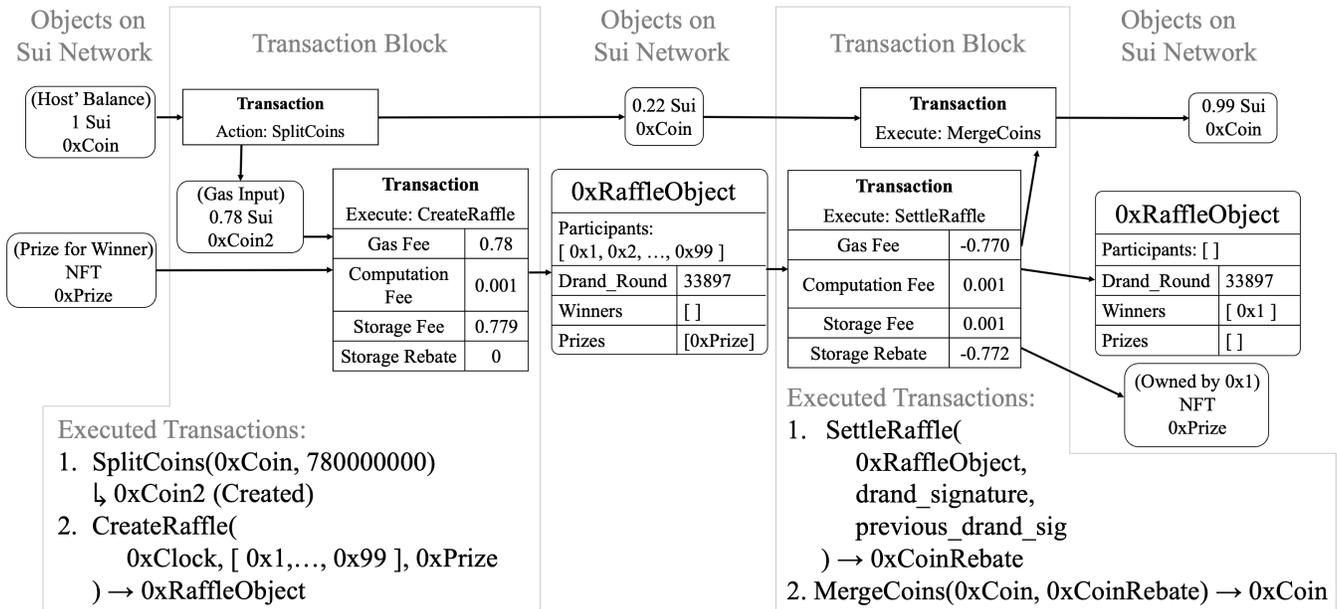

**Figure 1: Example of how Basic Raffle executed on the Sui Network. The first transaction block executed transactions to initiate the basic raffle via a DRAND round, while the second block settled the raffle with DRAND beacons and sent prizes to winners.**

However, there are three **limitations** to using the block's hash as a random value. Firstly, miners may intentionally calculate hashes that meet randomness conditions if the reward for generating a hash that meets certain conditions exceeds the block generation reward [2]. Secondly, the concentration of mining power in a few groups in today's PoW blockchain makes it susceptible to manipulation. Lastly, this method only applies to PoW blockchains, and as blockchain technology shifts towards more efficient methods of block production like Proof of Stake [10], in which a group of miners who are stakeholders takes turns producing blocks, block hash randomness is no longer applicable, as there is greater freedom for miners to manipulate block hashes when producing blocks.

### 3.2 Oracle-based Randomness

Committee-based Oracles are another common randomization[4]. Provided by third-party services, Oracles offer verifiable randomness for smart contracts, guaranteeing fairness in applications such as gaming and lotteries. When using Oracle-based random beacons, users first initiate a transaction and send it to the Oracle provider(s). Then, that provider sends a second transaction to input a committee-derived random value into the contract.

However, an Oracle-based random beacon has two **limitations**. Firstly, its usage requires costly fees (i.e. 3 USD per request on Ethereum). Moreover, Oracle lacks the decentralization levels of a Layer 1 blockchain, typically due to smaller committees and with different incentives or byzantine fault tolerance guarantees. If the

Oracle provider experiences downtime or is compromised by hackers, it may result in service interruptions and potentially impact the integrity of the service. There is also another potential issue related to long range attacks and the fact that the Oracles can perform blind, untraceable attacks (i.e. produce a future beacon before anyone else knows, and hence win a lottery without leaving footprints) [5]. It is necessary to have a more cost-effective and decentralized approach to ensure the availability of randomness on a blockchain DApp.

### 3.3 DRAND

Some modern blockchains support functionality to verify random beacons from the League of Entropy's DRAND, a distributed randomness beacon daemon [19]. The League of Entropy is a collaborative project that provides a verifiable, decentralized source of randomness accessible to anyone who needs public randomness. League members maintained the DRAND network by hosting nodes running the DRAND protocol. Note that although the DRAND network is also considered an Oracle service, we distinguish between paid Oracles, and a community transparent service like DRAND, cause use of it does not impose any paid services to the Oracle provider. Additionally, DRAND follows a standard cadence in producing publicly verifiable beacons, and it's not per request, hence reused by everyone for free.

The current DRAND mainnet network generates a random value every 30 seconds, called a "round"[1]. The generation process of a DRAND random value follows a specific set of steps [20]:

---

[1]As of Aug 3rd 2023, DRAND is also running a second mainnet with a 3-second random beacon cadence, and official developer docs are expected [22].

(1) **DRAND Network Setup**: To establish the network, each node is equipped with an identical private key by *Distributed Private Key Generation* [9] and agreed with a threshold. The threshold, typically set at 50% of the total node count, represents the minimum number of node signatures needed to reconstruct a complete signature.

(2) **Partial Beacon Creation and Broadcast**: For each new round, a node generates a *partial signature* by the current round number and the signature from the previous round. Then, broadcast it to the DRAND Network.

(3) **Final Signature Creation**: For each incoming *partial signature*, a DRAND node first verifies it and then stores it in a temporary cache if it is valid. Once a minimum threshold of valid *partial signature* is reached, the node combines them according to the BLS12-381 protocol [8]. This combination involves mapping these partial signatures to points on an elliptic curve, producing the final signature. By design, BLS signatures are deterministic; hence the threshold-combined BLS signature bytes can be considered a VRF output.

(4) **Validation and Storage**: The node validates the new signature by public key and previous signatures, then saves the signature with the round number to the database.

The random value of a DRAND round can be obtained by hashing the signature of that round with SHA-256. When utilizing DRAND for generating randomness within a smart contract, it typically involves the following steps:

(1) **Set Up**: When deployed, hardcode the threshold-combined public key of the DRAND network into the contract as a constant.

(2) **Initiate**: Calculate the current DRAND round number and designate a future round with a number greater than the current one as the target random value. Then, write the logic in the smart contract that should be executed based on this random value.

(3) **Verify and Execute**: Wait until the future round is announced at DRAND, then input DRAND's signature to the contract. The contract will first verify the legitimacy of the DRAND signature through the round number, public key, and the signature of the previous round. Following successful verification, it proceeds to hash the signature to derive the random beacon, subsequently executing the predefined logic with it.

The **advantages** of DRAND are that it is free, well recognized, decentralized, and has been widely adopted by many projects [19]. Additionally, thanks to the open nature of the DRAND Protocol, anyone can retrieve signatures from a DRAND Node and use them to invoke smart contracts for verification and execution once the designated DRAND round time passed.

However, there is a fatal **limitation** when using DRAND randomness: users need to wait at least 30 seconds to obtain results. Such a lengthy wait may influence the user experience in games.

## 3.4 Single-key VRF beacons

To enhance the immediacy of random outcomes, a more centralized but timely, unpredictable, and verifiable random technique has emerged, using a single entity VRF variation of the known commit-reveal schemes [3], which involves the following procedure:

(1) **Set up**: Write the host's public key as a constant into the smart contract.

(2) **Initiate**: The user inputs a seed into the smart contract. Then, the smart contract saves it with a unique and uncontrollable variable, such as a timestamp or block hash (this is to avoid replay attacks).

(3) **Sign the seed**: The host uses a private key to sign the seed and the unique variable at the backend and obtains the VRF output (i.e., BLS signature).

(4) **Verify and Execute**: The host calls the smart contract with the VRF output as a parameter. The smart contract verifies the validity of the signature using the host's public key, seed, and unique variable. Then, after successful verification, the smart contract hashes the signature to obtain a random value and execute the random logic.

The **advantage** of single-key VRF is its remarkable immediacy. After users initiate a transaction, the host can promptly sign and invoke the contract to complete the randomness. In products currently employing single-key VRF, users can receive the random result within five seconds after sending the initiated transaction, demonstrating low latency.

However, this method has two notable **limitations**. First, if a hacker compromised the VRF's private key, it could jeopardize the randomness, potentially leading to the complete depletion of the contract's assets. Moreover, only the host can *Verify and Execute*, as no one else can use the private key to create the signature / beacon. As a result, if the host encounters downtime, users may be unable to finish the execution and withdraw their assets from the contract. To circumvent the latter, solutions using time-locks have been proposed, by introducing penalties to VRF hosts who delay publishing their VRF outputs [5, 21].

## 4 SYSTEM DEVELOPMENT

In this section, we will begin by describing the implementation of the most basic raffle system, explaining its limitations, and introducing solutions. Then, we will gradually iterate our design to an advanced and production-ready raffle system. Source codes of different designs are at https://github.com/Bucket-Protocol/raffle-paper.

### 4.1 Basic Raffle System with DRAND

The goal of *Basic Raffle* is to enable the host to randomly select a winner from a group of participants and send the prize to that winner. The design of the *Basic Raffle* is intuitive and straightforward.

As shown in Figure 1, to initiate the random raffle, the host calls the Move smart contract's *create_coin_raffle*, which includes three key parameters: *Clock*, *Participants* and *Prize*. *Clock* is an object on the Sui platform that allows smart contracts to obtain the current time. *Participants* is an array containing all participants' addresses. *Prize* is an object about the reward for the raffle winner.

The *create_coin_raffle* function then performs the following:

(1) **Calculate the DRAND round**: The function calculates the current DRAND round according to the current time.

(2) **Initiate**: The function creates a *Raffle Object*. Raffle Object includes an array of participants, "*current DRAND round*

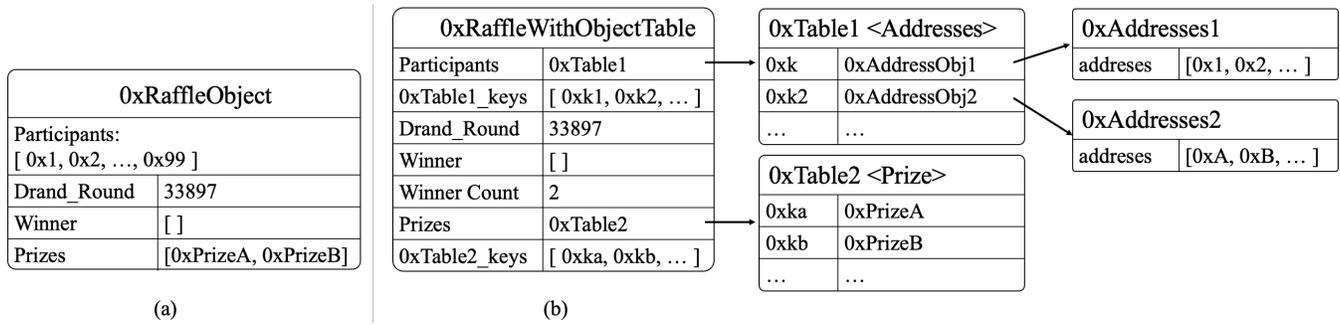

**Figure 2: Comparison between two raffle designs: one without an object table (a) and the other with a 1-layer object table (b). The design without an object table (a) on the left can hold up to $7,500$ addresses, whereas the design with a 1-layer object table (b) on the right can accommodate $7,500^2$ addresses.**

+ $N$" as the target DRAND round and the prize. N can be set to any number greater than 2, depending on when the creator wants to know the results.

(3) **Lock the prize**: If the prize is a fungible coin, the function converts the prize object to balance and saves it in a field of the Raffle. If the prize is not fungible, the function saves the prize as a sub-object.

Then, when the signature of the *current DRAND round + N* is revealed, anyone can settle the raffle by calling *settle_coin_raffle*, which perform the following steps:

(1) **Verify**: The function checks the input DRAND signature is valid.

(2) **Execute**: The function first computes a random value based on the DRAND signature and subsequently uses this random value to pick a winner from the participants' array. It then proceeds to transfer the prize to the selected winner.

(3) **Settle**: The function emits events about the raffle result. Then, clear the participants' array to release space.

The creation and settlement process is illustrated at the first and second *Transaction Block* in Figure 1.

### 4.1.1 Advantages.
The *Basic Raffle* has three advantages. Firstly, it fully utilizes DRAND's capabilities, ensuring that the Raffle's outcome remains unknown initially while allowing anyone to settle and verify the results. Additionally, the rewards are securely locked within the contract once the Raffle is created, preventing organizers from running multiple Raffles and picking the favorable result, thereby guaranteeing fairness in the process. Lastly, the settlement process involves data clearance and offers a *Storage Rebate*, creating a strong incentive for anyone to settle the raffle.

### 4.1.2 Limitations.
However, there are several limitations when considering Basic Raffle for production use. First, Sui limits the size of transaction blocks, permitting an array size input of approximately 400 addresses. Consequently, the creation process will encounter errors if the host intends to include more than 400 participants in the Raffle. Furthermore, there is a size restriction of 250 KB for Sui Objects. Even if the host manages to input a sufficient number of addresses, the Raffle object cannot accommodate more than 7500 addresses. Therefore, it becomes necessary to leverage other features of Sui to accommodate more addresses.

## 4.2 Raffle with Object Table
The *Object Table* feature within Sui enables a parent object to own other objects. functioning in a manner akin to *Mapping* in *Solidity* or a *Dictionary* in *Python*. As depicted in Figure 2, *Object Table* empowers a *Raffle Object* to possess multiple sub-objects, such as several *Prize Objects* and *Participant Address Objects*.

Under these circumstances, we can establish a mechanism that automatically splits participants and prize addresses into separate sub-objects according to size. When the size surpasses 7,500, a fresh sub-object can be created and put under the *Object Table*, with its corresponding key added to the *Object Table* key list. In the case of using a single-tier *Object Table*, this strategic setup allows a Raffle to accommodate an impressive $7,500^2$, which equals $56,250,000$ addresses, a capacity more than sufficient for most practical scenarios. Additionally, it is possible to implement multiple levels of *Object Table* if even larger quantities are needed.

### 4.2.1 Advantages.
Through the use of *Object Table*, we have addressed the issue of the number of addresses that an Object can accommodate. This type of Raffle is suitable for situations where users actively join by paying for themselves, such as a lottery. An example scenario is as follows:

(1) **Initiate**: The host initiates the raffle by locking the prizes within the contract, setting the entry fee, and establishing a deadline using DRAND's Round.

(2) **Participants Join**: Participants submit their entry fees to the contract. The smart contract verifies that the deadline has not been exceeded and the fees are valid. Then, it adds the users' address to the *Address Object* within the *Object Table* fields of the *Raffle Object*. If the *Address Object* is full, a new *Address Object* is created under the *Object Table*.

(3) **Settle**: When the designated time arrives, anyone can settle using DRAND Signatures, and the prizes are sent to the winner's wallet while the entry fees are transferred to the host's wallet.

### 4.2.2 Limitations.
In the above case of the raffle, the block size won't be an issue as users enter one address at a time. However, if we wish to input many addresses at once, we still have to deal with the block size limitation of only being able to enter 400 addresses in a single attempt. This limitation becomes particularly cumbersome when, for example, users who want to create a raffle with 2000 addresses must go through the signing process five times. This problem can be resolved using the *Delegate Object Creation* or *Zero-Knowledge Proof* approach.

### 4.3 Raffle with Delegate Object Creation

*Delegate Object Creation* aims to streamline the process for users with the help of a *Delegate Host*, enabling them to efficiently input numerous addresses into *Raffle Object* at the Sui Network with just one wallet operation. The steps of Delegate Object Creation are depicted in Figure 3 and described in the following:

(1) When a user wants to create a Raffle, he first sends all addresses to the backend of the Delegate Host.

(2) The Delegate Host then batches these addresses and writes them into an *Delegated Object* using its backend's Private Key while paying the storage fee. The fee will reflected as a usage fee in the field of *Delegated Object*.

(3) Then, the user can initiate the raffle with the address table in *Delegated Object* in a single transaction. Moreover, when using the Address Object, the user pays the storage fee and transaction fees to the Delegate Host.

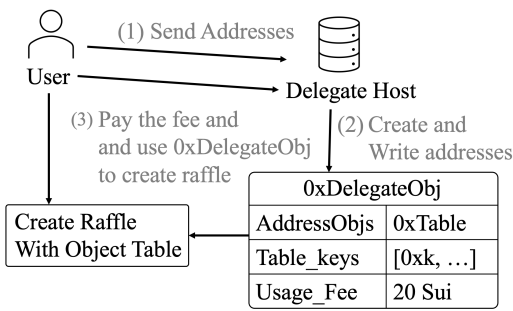

**Figure 3: Flows of how Delegate Object Creation works.**

When users settle a raffle, they can rebate the storage fee by clearing their data to reclaim most usage fees. Additionally, if users do not use the Delegated Object for certain times, the Delegate Host can proactively remove data from the Delegated Object to rebate the storage fee.

*4.3.1* **Advantages**. The advantage of Delegate Object Creation is that users only need to open their wallet and perform the signature once to create a Raffle with lots of participant addresses. By doing so, all participants' addresses are stored on the transaction history of Sui Network, allowing every participant to verify their inclusion in the Raffle.

*4.3.2* **Limitations**. The limitation of *Delegate Object Creation* is the waiting time. At the backend, it typically takes 1 second per transaction to upload 400 addresses. If there are 20,000 addresses, users will experience approximately a 50-second waiting time. Moreover, if users only need to select a few winners from many addresses, this method of writing and clearing addresses is inefficient. Consequently, we have developed another approach using Zero-Knowledge Proofs to quickly create a raffle with many addresses.

### 4.4 Raffle with Zero-Knowledge Proof

Zero-knowledge proof (ZKP) is a cryptography and mathematical technique [13] employed within the realm of blockchain to verify the authenticity of a statement while preserving the confidentiality of the statement's details. The uses of ZKP in blockchain now extend beyond safeguarding privacy [24] but also encompass optimization in on-chain data storage to enhance efficiency [11]. The most renowned use of ZKP for storage optimization is Ethereum's Zero-Knowledge rollups (ZK-rollups) [11] with Merkle Tree.

Merkle Tree is a cryptography binary tree structure used for data validation [15]. Merkle Tree ensures data integrity and facilitates efficient verification, making it particularly advantageous when handling extensive data while optimizing bandwidth and computational resources. Illustrated in the right side of Figure 4, in Merkle Tree, each leaf node corresponds to a data block within this structure, whereas each non-leaf node stores the hash of its child nodes. The uppermost node above all the leaves is called the Merkle Root. The number of data blocks that a Merkle Tree can accommodate is related to its number of layers. A Merkle Tree with $N$ layers can accommodate $2^N$ data blocks.

To confirm the authenticity of a target data leaf, one only needs to calculate the hash of the target leaf and hash it with the hashes of intermediate Merkle nodes (referred to as the Merkle Proof), then compare the outcome with the Merkle Root to ensure they match.

With the implementation of Merkle Tree, ZK-rollups can handle many transactions with a low transaction fee as it only publishes the Merkle Root of these transactions to the Mainnet. A similar mechanism, as illustrated in Figure 4, can also be applied to optimize data storage for the raffle system. The steps for using Merkle Tree to handle participants' addresses in a ZK-raffle include the following:

(1) **Prepare**: Collect and organize the addresses of all participants.

(2) **Generate**: Generate the Merkle Tree and store it in the backend database.

(3) **Initiate**: Create the ZK-raffle by incorporating the Merkle root, prizes, and the count of Merkle leaves. The smart contract then calculates the DRAND round based on the current time.

(4) **Settle Raffle**: Employ the DRAND Signature to determine the raffle's outcome and identify the list of IDs for winners.

(5) **Claim Reward**: Anyone can use the ID, participant's address, and Merkle Proofs to claim the prize. The smart contract will check the ID is in the winner list and that the hash of the ID and address with Merkle Proofs is aligned with the Merkle root. After claiming, the ID will be removed from the winner array to avoid double-claiming.

*4.4.1* **Advantages**. The ZK-raffle offers four notable advantages. Firstly, thanks to the lightweight nature of ZKP, the host can launch a ZK-raffle with lots of participants with low transaction fees since the host only needs to store the Merkle Root on the smart contract to represent numerous addresses. Additionally, because the Merkle Tree generation takes place in the backend, the ZK-raffle can be computed and prepared in an extremely short amount of time. Furthermore, this method eliminates the need for Storage Rebates, reducing the financial burden for the raffle creation and the load on the network.

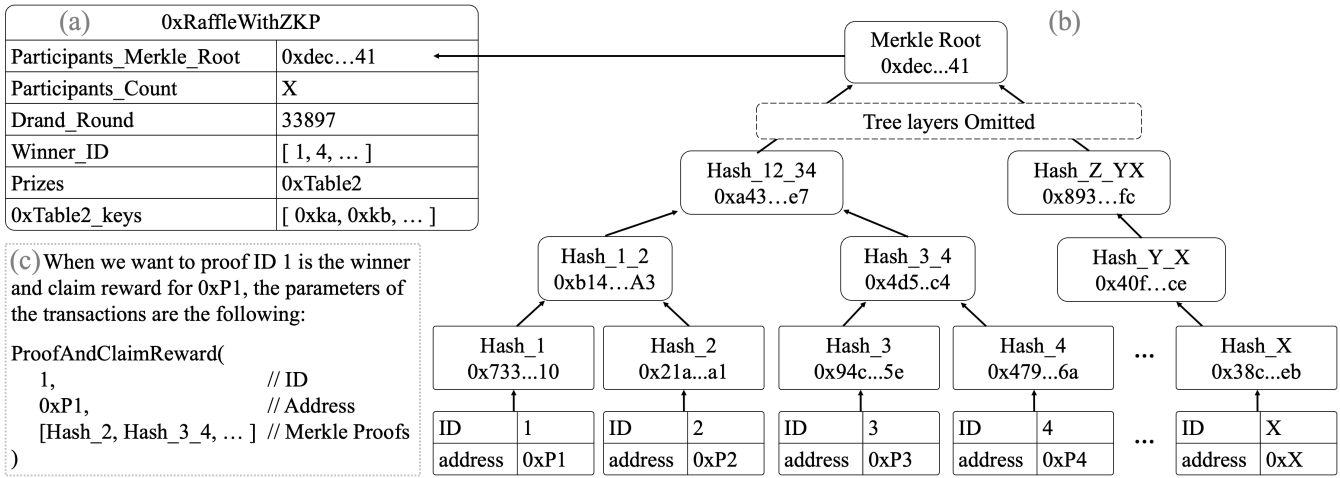

**Figure 4: Example of Raffle with Zero Knowledge Proof. The left (a) is the Raffle Object. The right (b) is the Merkle Tree computed with raffle participants. At left bottom, (c) is the example parameters to proof and claim reward with Merkle Proofs.**

*4.4.2 Limitations.* ZK-raffle has two limitations. First is its complicated prize distribution process. Each prize distribution involves multiple rounds of hashing calculations with Merkle Proofs, which can be costly when there are many winners. Second, the Merkle Tree's storage tends to be centralized with a database containing all Merkle Tree and Proofs data. This database is essential for users to confirm and claim their rewards or verify their raffle participation. If the database encounters data custody issues resulting in the loss of Merkle Proofs, all rewards may become locked within the smart contract.

## 4.5 ZK-raffle with single-key VRF

In addition to DRAND Randomness, employing a ZK-raffle with the host's VRF outputs is a wise choice because they share common limitations, necessitating a centralized host to provide Merkle Proof or single-key VRF to settle the raffle. Moreover, since the Merkle root won't be cleared after the raffle is settled, it is possible to use the same raffle multiple times. Therefore, they can be integrated to create a fair web3 raffle system, allowing users to draw random prizes fairly.

Illustrated in Figure 5, implementing a ZK-raffle with Signature Randomness includes the following steps:

(1) The host initiates the ZK-raffle smart contract, providing essential parameters such as the Merkle Root, prize count, and public key, while setting the entrance fee at 10 Sui.

(2) A user enters the raffle by paying the 10 Sui entrance fee. Subsequently, the raffle smart contract generates a raffle ticket that includes the fee and forwards it to the host.

(3) The host settles the raffle ticket by verifying the VRF output (i.e., BLS signature) and computing a random result through the signature. Afterward, by validating the Merkle Proof to confirm that the random result indeed corresponds to the Prize NFT provided by the host, the host received the 10 Sui Entrance Fee while the user received the Prize NFT.

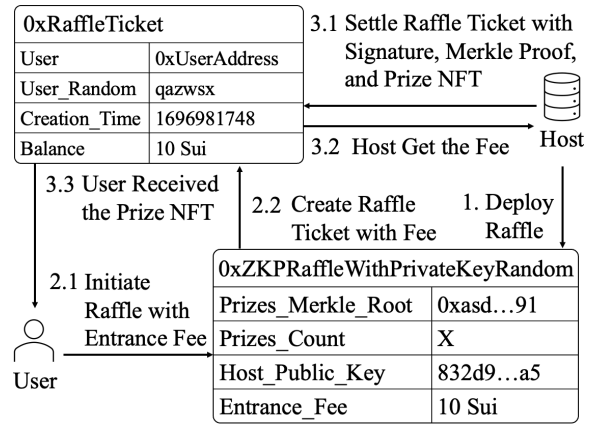

**Figure 5: Example flows of how Zero Knowledge Raffle with host's VRF (signature) works.**

*4.5.1 Advantages.* A ZK-raffle with host's VRF offers three significant advantages. Firstly, it is highly lightweight, avoiding data storage issues commonly associated with other versions of raffles. Secondly, it ensures that the host cannot manipulate the probabilities, as anyone can calculate the Merkle Proof to determine their chances of winning the overall prize, making it ideal for prize raffles such as gaming loot-box. Thirdly, it strikes a good balance between centralization and decentralization. While both Merkle Proofs and Signatures require centralized resources, the host must provide these resources to claim the entrance fees from users. Consequently, the host is strongly incentivized to settle the raffle.

*4.5.2 Limitations.* On the other hand, there exist two limitations. Firstly, since the randomness is generated from a single private key, the owner of that key (typically the host) can produce beacons at their discretion, granting them the ability to manipulate the raffle result in their favor. In the hands of a malicious host or a host compromised by hackers, this private key can be used to exploit this vulnerability, resulting in the drawing of numerous grand prizes and undermining the distribution of prizes. Secondly, if the host refuses to settle the raffle, the users' entrance fee remains locked within the contract. Nevertheless, this limitation can be mitigated by leveraging the transparent nature of blockchain. We can ensure that the host acts fairly by monitoring the number of unsettled raffle tickets and even giving penalties accordingly. For instance, if the host deliberately avoids settling a particular raffle ticket, it may indicate dishonest behavior, and users can claim their winning once a predetermined time period has passed.

| Randomness Type | Host's VRF Randomness | DRAND Randomness | | |
|---|---|---|---|---|
| Raffle Type | Basic Raffle | Basic Raffle | Raffle with Object Tables | ZK-Raffle |
| **Initiate Raffle Fee with 200 participants** | 66 | 46 | 454 | 5 |
| **Settle Raffle Fee with 10 winners** | 67 | 86 | 234 | 599 |
| **Settle Fee increased per winner added** | 6 | 6 | 9 | 59 |
| **Total Transaction Fee** | 133 | 132 | 688 | 604 |
| **Total Transaction Fee in Sui** | 0.00009975 | 0.00009900 | 0.00051600 | 0.00042225 |
| **Total Transaction Fee in USD** | 0.0000408975 | 0.0000405900 | 0.0002115600 | 0.0001731225 |

Table 1: Comparison of transaction fees for various randomness approaches and raffle system designs. These fees are assessed based on the computational costs and do not include storage fees. The Fee in Sui is calculated according to the recommended Gas Price on Sui with formula $Fee * 750 * 10^{-9}$. The Fee in USD is calculated according to the Sui price of $0.41$ USD on 2023/10/12.

## 5 TRANSACTION FEE COMPARISON

To compare the transaction fees of different Raffle designs, we developed minimum viable smart contract codes and used the testing functionality of Sui Move Cli to obtain the transaction fee of each design. The source code is available at https://github.com/Bucket-Protocol/raffle-paper.

The transaction fee of a raffle involving 200 participants and resulting in 10 winners is displayed in Table 1. The findings show that transaction fees are consistently cheap regardless of the method used. The results demonstrate that the transaction fees for *Signature Randomness* and *DRAND Randomness* are roughly equivalent. Additionally, more advanced raffle designs incur higher transaction fees. The setup fee for the raffle with *Object Table* is more expensive due to the process of *Object Tables*. Regarding *ZK-Raffle*, the settlement cost is just 8, but each winner must spend a computation fee of 59 to validate the *Merkle Proof*, making it more costly overall.

## 6 DISCUSSION

Through the case study of building a random, fair, and verifiable raffle game on Sui Network, we explored various design approaches, inherent limitations, and potential solutions when developing games with randomness using smart contracts. Each design option comes with its unique set of advantages and drawbacks, and the choice should be made depending on the particular circumstances at hand.

### 6.1 Randomness on Blockchain

Two methods for achieving randomness in smart contracts are considered best practices: DRAND Randomness and host's key Signature Randomness. As shown in Table 1, both methods have a comparable transaction fee. Drand Randomness aggregates input from nodes across the DRAND network to generate random values, making it more decentralized and secure. However, the current official (before the mainstream update to a 3-second cadence) using DRAND requires a 30-second wait each round, which could impact the user experience. On the other hand, single-key VRF randomness generates signatures using the raffle host's private key for randomness and is faster. However, it is more centralized, requiring the host to secure the private key and provide a signature each time.

### 6.2 Smart Contract Design Limitations, Solutions, and Recommendations

When building a raffle in smart contracts, it's critical to consider network limitations, especially concerning the size of arrays and objects. We propose two solutions to address size constraints: "Raffle with Object Tables" and "Zero Knowledge Raffle."

"Raffle with Object Tables" involves transforming the content of arrays into a multi-layered table data storage structure, maximizing the data storage capacity. The advantage of this approach is that all data remains on the blockchain. However, it has higher computational costs, requires multiple setup transactions, and incurs higher storage fees. The need for multiple setup transactions can be mitigated using the "Delegate Object Creation" method, where a delegate host assists in creating Object Tables. Storage fees can be reduced by reclaiming storage space through Storage Rebates by clearing data at the end of the Raffle.

The raffle with Object Tables is perfect for situations that require absolute transparency and decentralization, such as selecting lucky winners among players to receive exclusive event prizes. Additionally, a host can utilize Raffle with Object Tables to create games like lotteries that require active participation from players.

On the other hand, the Zero Knowledge Raffle transforms array fields into a Merkle Tree, enabling an infinite capacity of addresses by uploading only the root of the Merkle Tree to the smart contract. However, it comes with higher transaction fees when verifying the Merkle Tree during prize distribution.

The Zero Knowledge Raffle is well-suited to reduce transaction costs in scenarios involving many candidate entities, such as Loot Box and Gacha Games. Moreover, by combining Private Key Randomness with the Zero Knowledge Raffle, it is possible to enhance the efficiency of the raffle while ensuring decentralization and incentives for the host to provide a Merkle Proof to settle the raffle.

# 7 CONCLUSION

This paper explained how to build a fair, verifiable, efficient random raffle on the Sui Network. We investigate a range of smart contract design methodologies for constructing chance-based games through smart contracts, identify their inherent constraints, and propose remedies. Furthermore, we furnish insights drawing from the advantages and limitations of different design choices. Through these insights, we provide valuable guidance to researchers and developers engaged in randomness games using smart contracts.